\documentclass[12pt]{article}

\usepackage[pdftex]{graphicx}
\usepackage{epstopdf}
\usepackage{stackrel}
\usepackage{amsmath,amssymb,exscale}
\usepackage{array,multicol}
\usepackage{afterpage,float,flafter}
\usepackage{cite}
\usepackage{color}
\usepackage{adjustbox}
\usepackage{wrapfig}
\usepackage{caption}

\@addtoreset{equation}{section} \makeatother

\providecommand{\norm}[1]{\lVert#1\rVert}

\begin{document}

\title{\hfill {\small \medskip }\\
\textbf{Formation of spiky structures in
high-altitude snow patches: penitente tilting}}
\author{Pablo Guilleminot$^{a}$ and Rodrigo Olea$^{a}$\smallskip \\
$^{a}${\small \emph{Departamento de Ciencias F\'{\i}sicas, Universidad
Andres Bello,}}\\
{\small \emph{Sazi\'e 2212, Piso 7, Santiago, Chile. \smallskip }}\\
{\small \texttt{yemheno@gmail.com, rodrigo.olea@unab.cl}}}
\date{}
\maketitle

\date{}
\author{}
\maketitle

\begin{abstract}

\emph{Penitentes} are spikes formed on the surface of the snow, which are present typically at high altitude
in the Andes and Himalayas. They are a consequence of a thermodynamic instability, as a result of the surface sublimation at a given point due to the incidence of light scattered by surrounding features.
Here, based on existing literature, we model the time evolution of penitente formation as a purely radiation-driven phenomenon.
The physical system is governed by a $1D$ diffusion equation with a nonlocal source term, which represents
the light coming in from all the line of sight accessible from that point of the curve. For small
perturbations on the initial profile, the surface undergoes an instability which
triggers the formation of spiky structures.
For solar radiation coming in the surface at a given angle, our numerical simulations account for a feature
observed in the real system: penitentes get tilted in the direction of the sunlight.
\end{abstract}

\thispagestyle{empty} \vspace*{.5cm}

\newpage \renewcommand{\thepage}{\arabic{page}} \setcounter{page}{1}

\section*{Introduction}

Thermodynamics is at the very root of the understanding of physical
phenomena, including the ones we observe in everyday life. Some of them
are so unusual and counter-intuitive that are a matter of study until
now.

\captionsetup[figure]{font=footnotesize}
\begin{wrapfigure}{r}{0.3\textwidth}
\includegraphics[width=0.3\textwidth]{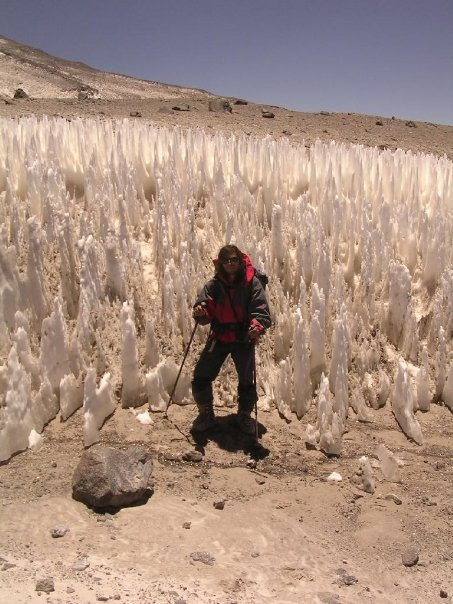}
\caption{\footnotesize{One of the authors (R.O) in a penitente field at 5.500 m.a.s.l. in volcano Ojos del Salado, Atacama, Chile.}} \label{RO}
\end{wrapfigure}
\captionsetup[figure]{font=footnotesize}

Under certain conditions, at high altitude, the surface of snow
develops spike-like structures known as \emph{penitentes}. This is a
consequence of surface sublimation at a given point as a result of both
direct sunlight and light scattered by the surrounding region \cite{Mat,RAW,Ams}.

Once the process of ablation starts, the
surface geometry provides a positive feedback mechanism, and
radiation is trapped by multiple reflections between the walls \cite{Bartels,Bergeron}.

Different factors may be considered when it comes to modeling the formation of structures on the surface of snow patches: surface reflectance, wind, humidity, angle of the light source, etc \cite{dust,Claudin}.

The key condition for an icy system to develop spike-like structures on its surface is that the dew point must be below the freezing point \cite{Lliboutry}. This implies that the rate of ice loss at the air/snow interface exceeds the capacity of capturing vapor from the surrounding air. This condition favors sublimation over surface melting, and it is usually met at high altitude in intermediate latitudes.  On the contrary, if the snow surface melted,
the water would weigh down and smooth out any spiky features.
This explains why penitentes are found in Cordillera
de Los Andes and Himalayas and, only occasionally, in other mountain ranges. The appearance of this phenomenon seems to be more common than expected, as recent literature has reported the presence of penitentes in Jupiter's moon Europa \cite{Eu}
and even in ice fields in the distant Pluto \cite{Pl}.

This article models the time evolution of penitente formation as an instability
of a diffusion equation that controls the surface sublimation at a given point,
with a source term given by the incident solar radiation. Thus, we  neglect the effect of the other sources of ice ablation mentioned above \cite{WW}.

\section{Toy model for the formation of surface spikes} \label{eq}

At a given point $x$ of the curve that defines the snow profile, the ablation is proportional to the heat absorbed due to
both direct sunlight or light scattered by the surrounding region. A part of the radiation is reflected and a part of the heat is lost by convection
to the air. The remaining fraction is transmitted to the neighboring points.
The heat transmission within the surface is governed by the diffusion equation \cite{Betterton},

\begin{equation} \label{main1}
\frac{\partial U}{\partial t}= \Upsilon(x)+D\frac{\partial^2 U}{\partial x^2}\,,
\end{equation}
\begin{equation} \label{main2}
\frac{\partial h}{\partial t}= -K\frac{\partial U}{\partial t}\,,
\end{equation}
\smallskip
\noindent
where the function $U(x,t)$ is the thermal energy on the point $x$ at a time $t$, $h(x,t)$ is
the profile height, $D\frac{\partial^2 U}{\partial x^2}$ is the diffusion term,
$K$ is a proportionality constant and $\Upsilon(x,t)$ is the energy
which acts as a source at $x$.

\subsection{Construction of the geometric model}

\begin{figure}[h]
\centering
\includegraphics[trim={3cm 8cm 3cm 8cm},clip,width=0.5\textwidth]{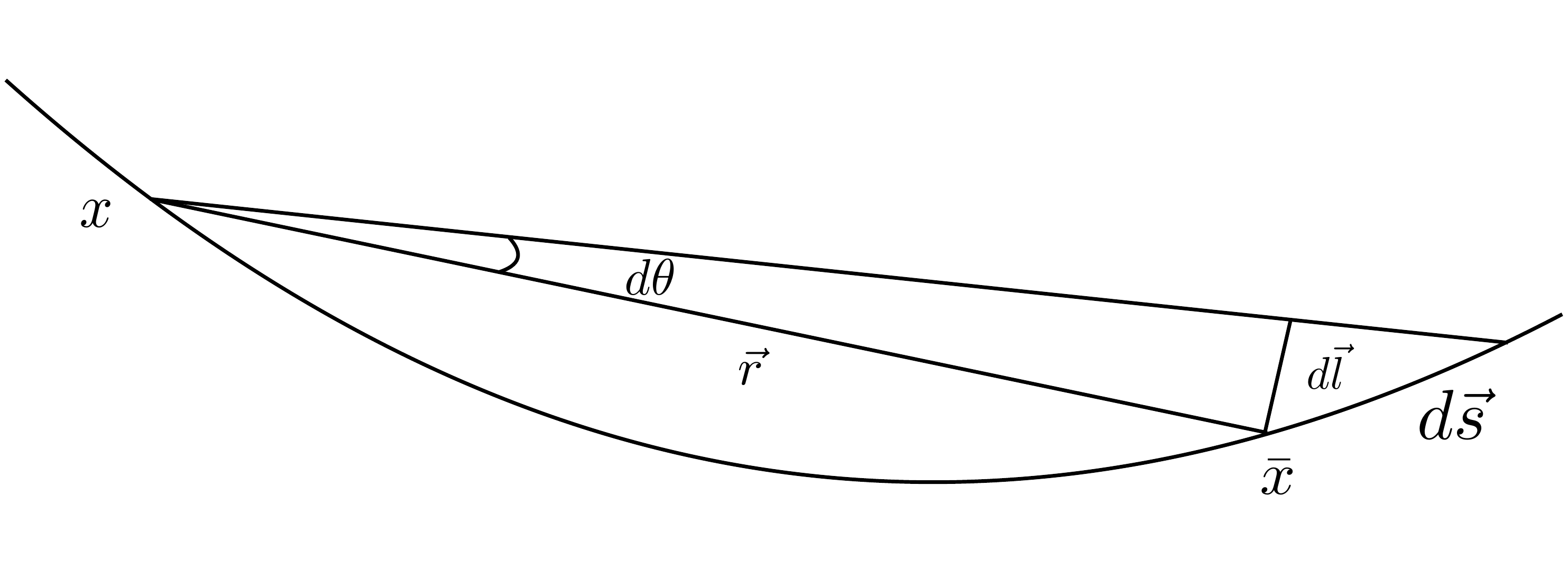}
\caption{Graphic illustration of snow surface} \label{fig1}
\end{figure}

Snow surface looks white because of a diffused reflection
of the incident sunlight. As the light is partially reflected beneath the surface,
all colors are scattered roughly equally in all directions.
Due to this fact, every single point of the curve acts as a source of equal intensity per
unit of length, which is spread out over the line tangent to that point. Neglecting  tiny differences in reflection along the curve means that
the surface geometry is all that matters when one computes reflection patterns on a
particular point of the curve.
For an infinitesimal interval of length $ds$ in the reflecting surface
(at a point $\bar{x}$), the contribution  to the absorbed energy at another point $x$ is
proportional to the sustained angle $d\theta$ \cite{Betterton} (See figure \ref{fig1}).

The function $\Upsilon(x)$ considers the absorption from direct sunlight and subsequent reflections. Up to
the first light bounce, this function is given by \cite{Betterton}
\begin{equation} \label{U1}
 \Upsilon(x) = I_0\alpha\left [ 1+(1-\alpha)\Xi(x)\right ]\,,
\end{equation}
where $I_0$ is the energy of sunlight and $\Xi(x)$ stands for the fraction of energy received by neighbouring portion of the surface.

As suggested by the figure, $\Xi(x)$ comes as the sum of all the
infinitesimal contributions coming from the \emph{line of sight} (LoS)
\begin{equation}
\Xi(x)=\frac{1}{\pi}\underset{LoS}{\int}d\theta =\frac{1}{\pi}\underset{LoS}{\int}\frac{\norm{r\times ds}}{r^2}%
=\frac{1}{\pi}\underset{LoS}{\int} \frac{|h'(\bar{x})\Delta \bar{x} -\Delta h|}{(\Delta \bar{x})^2+(\Delta h)^2}d\bar{x}\,,
\end{equation}
\noindent
where $\Delta \bar{x}$ and $\Delta h$ are given by $\Delta \bar{x}=|x-\bar{x}|$ and $\Delta h =|h(x)-h(\bar{x})|$, respectively.

The arc element seen by an observer at $x$, $d\theta$, can be expressed in terms of $x$ and $h(x)$ from the above figure. It is simply given by $d\theta=\frac{dl}{r}$, where $dl$ is the
height of the parallelogram formed by $\vec{r}$ and $d\vec{s}$.

\subsection{Line of sight}

Let us consider an arbitrary point $P\left(x,h(x)\right)$ on the curve. For that point, the line of sight
is defined as the collection of points $Q\left(\bar{x},h(\bar{x})\right)$ with $\bar{x}\in[a_i,b_i]$
that is connected by a straight line without intersecting the surface. The extreme of the interval $b_i$ is given by the tangent
line to the snow profile, i.e., the condition

\begin{equation}
h^{\prime }(b_i)=\frac{h(b_i)-h(x)}{b_i-x}\,.
\end{equation}

\begin{figure}[H]
\centering
{\includegraphics[width=0.5\textwidth,trim=5cm 8cm 3cm 5cm,clip]{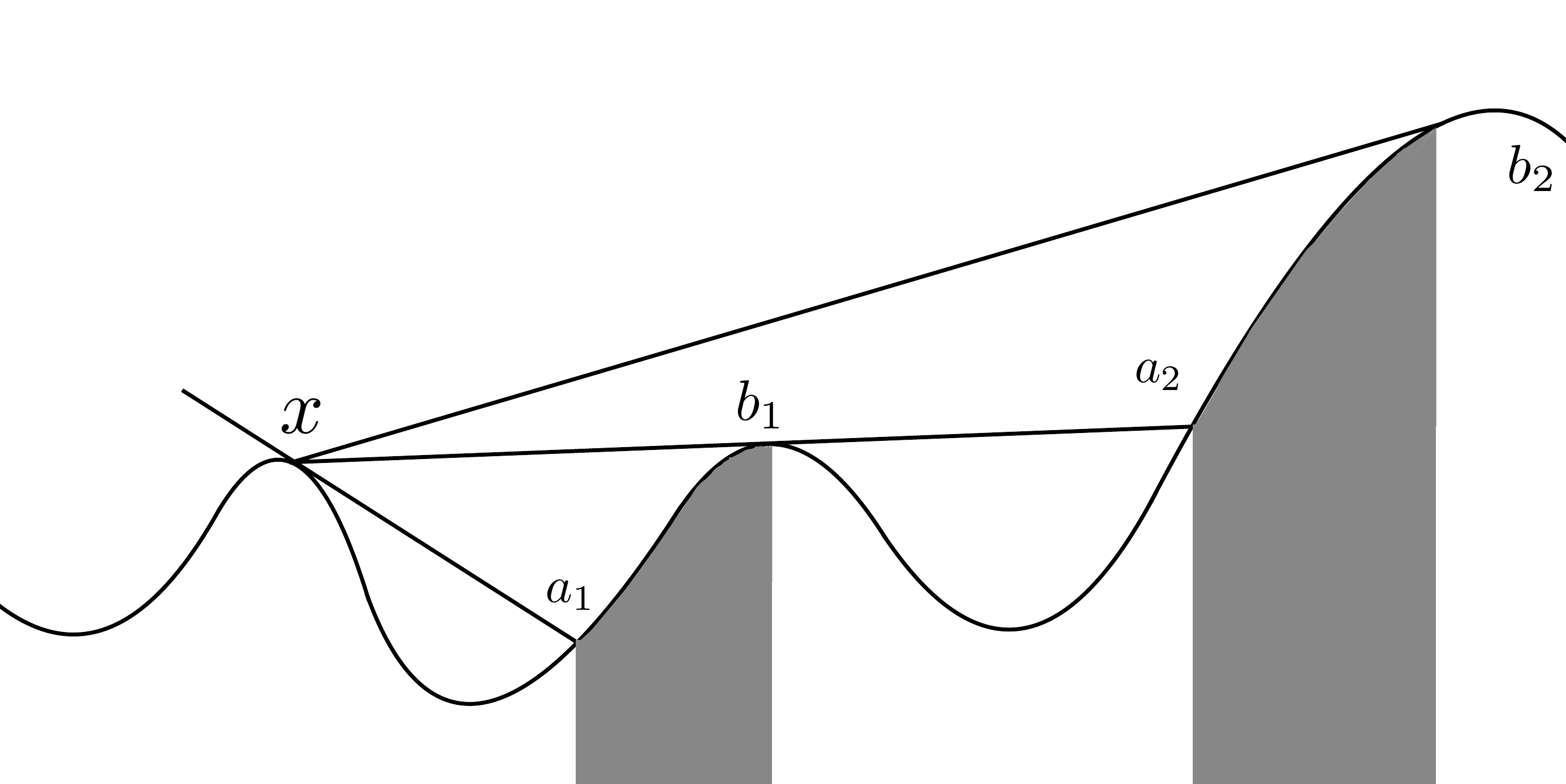}}
\caption{Line of sight} \label{fig2}
\end{figure}

The surface must be locally convex, i.e, $h''(b_i)<0$, in order to
exclude points at a valley. The condition for $a_i$ is obtained from
the extension of the line that connects $P$ with $b_{i-1}$
(See figure \ref{fig2}), such that

\begin{equation}
h(a_i)=h^\prime(b_{i-1})(a_i-b_{i-1})+h(b_{i-1})\,.
\end{equation}

The above relations define a set of piecewise intervals that determine
the integration domain of $\Xi(x)$.

\subsection{Shady zones} \label{shady}

Commonly the sunlight does not come in on the snow surface in a vertical line, what explains the \textit{in situ} 
observation that penitentes tilt in the direction of the sun \cite{Mat}.

If the solar radiation falls upon the surface at a given angle $\phi$, we need to exclude the contribution to $\Xi(x)$ from the  points in the shadows produced
by the peaks. Geometric considerations indicate that the first point in the shadow,
namely $x=c$, must satify,
\begin{equation}
\label{eq:c}
h^\prime(c)=\tan(\phi)
\end{equation}
Also, the profile curve must be locally convex at $x=c$. Assuming that the light is coming from the left,
the points in the shadow are below the tangent line to the point defined above, i.e.,
\begin{equation}
 h(x)\leq h^\prime(c)(x-c) + h(c)\,.
\end{equation}
We should stress that even tough the shady zones do not receive the first light ray, they eventually sublimate
due to scattered light. Finally, the source for the diffusion equation (\ref{main1}) is given by
\begin{equation} \label{asd}
 \Upsilon(x) = I_0\alpha\left [ \mathbb{I}(x)+(1-\alpha)\Xi(x)\right ]\,,
\end{equation}
where $\mathbb{I}(x)=1$ if $x$ is a lighted point and $0$ otherwise. This
expression takes into account contributions up to the first reflection on the surface.

\section{Methodology and results}

In order to analyze the time evolution of the system, the partial differential equations in Sec.\ref{eq} were solved numerically using a computer code developed in C++.
The initial surface profile was taken as either sinusoidal functions or the product of them, which resembles actual wind-generated microfeatures.

The code follows a standard finite difference method, with second-order central difference for the diffusive term and forward difference for the time. It also discretizes the function $h(x)$, in order to calculate the line of sight to obtain the value for the source $\Upsilon(x)$ numerically.
Eq.(\ref{main1}) is then used to determine the height of the snow/air interface at the point $x+\Delta x$.

Below, we display the main results
in the form of graphic plotting at different times, what corresponds to the evolution of the snow surface shape.

\begin{figure}[H] \label{prof1}
\center
    \fbox{\includegraphics[width=0.45\linewidth]{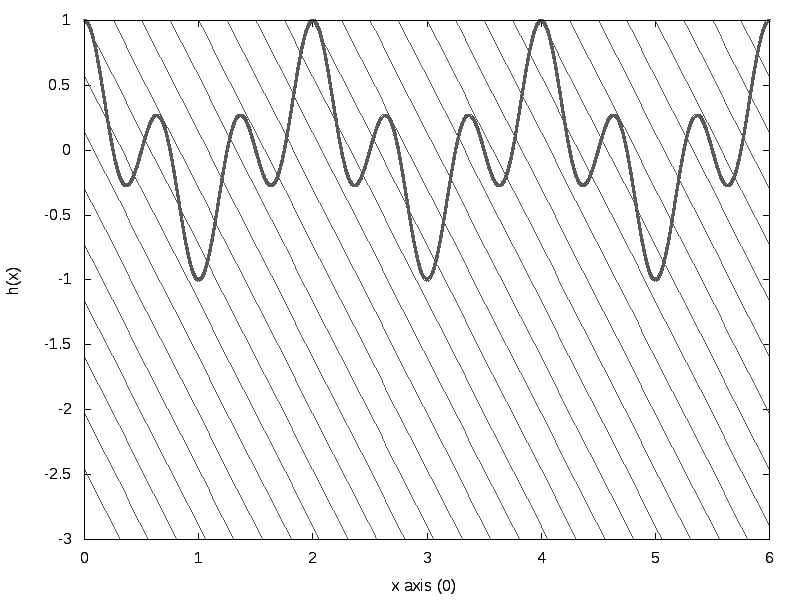}}
    \fbox{\includegraphics[width=0.45\linewidth]{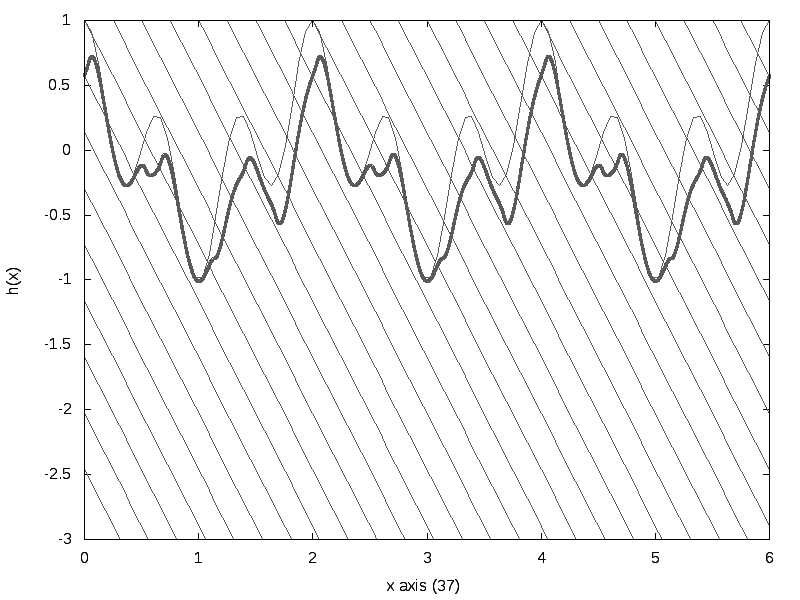}}
    \fbox{\includegraphics[width=0.45\linewidth]{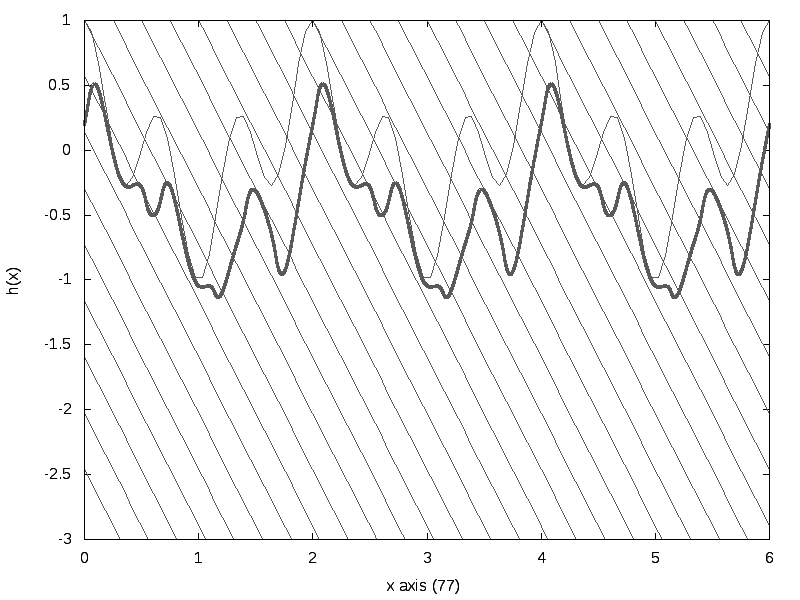}}
    \fbox{\includegraphics[width=0.45\linewidth]{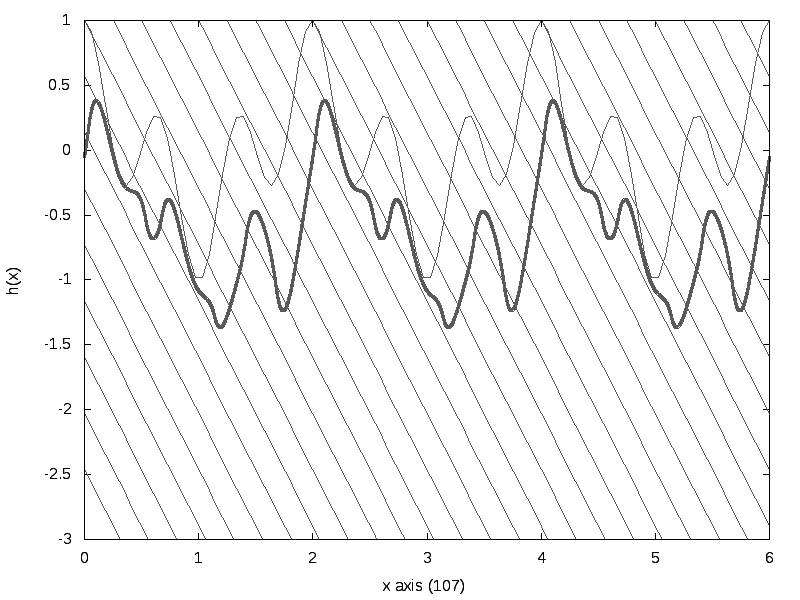}}
    \caption{Time evolution for $h(x)=\cos( \pi x)\cos(2 \pi x)$}
\end{figure}

In the previous sequence, the thin line marks the initial profile and the tilted lines shows the direction
of the incident sunlight. The thick line plots the function $h(x)$ for a given time. The values of the constants are $K=1$, $D=1$ and $\alpha=0.5$. It can be seen from that image that the structures try to align along the direction of the sunlight, producing the penitente tilting.

\begin{figure}[H] \label{prof2}
\center
    \fbox{\includegraphics[width=0.45\linewidth]{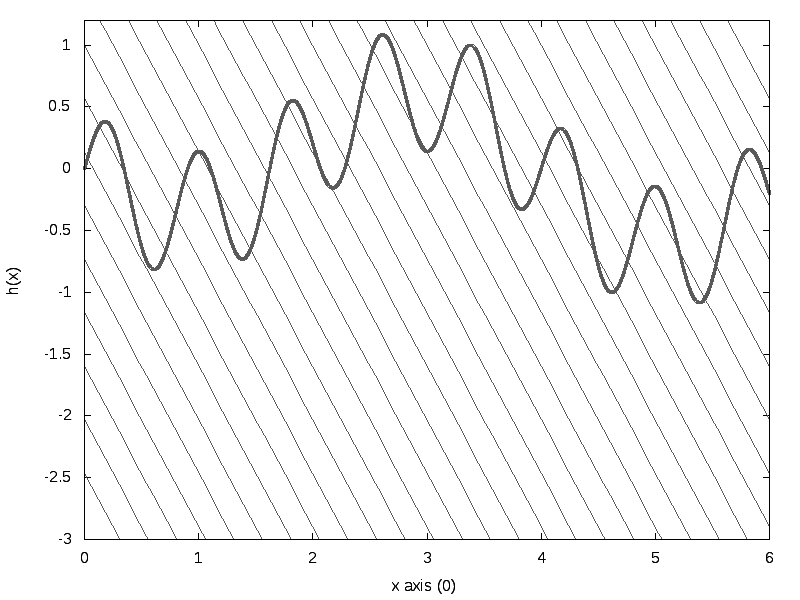}}
    \fbox{\includegraphics[width=0.45\linewidth]{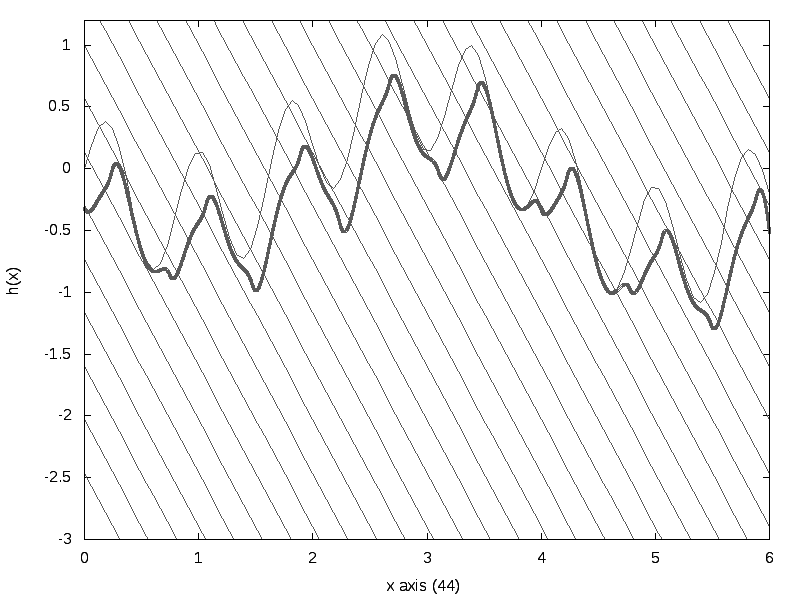}}
    \fbox{\includegraphics[width=0.45\linewidth]{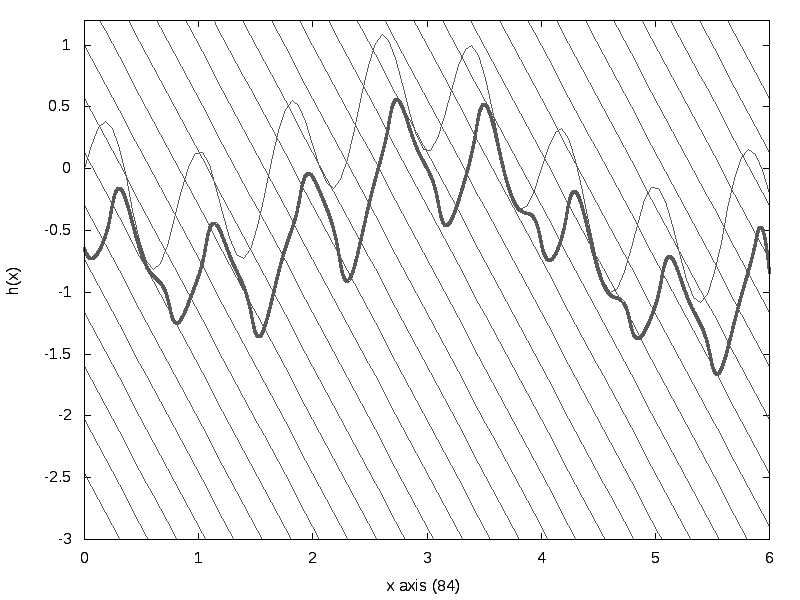}}
    \fbox{\includegraphics[width=0.45\linewidth]{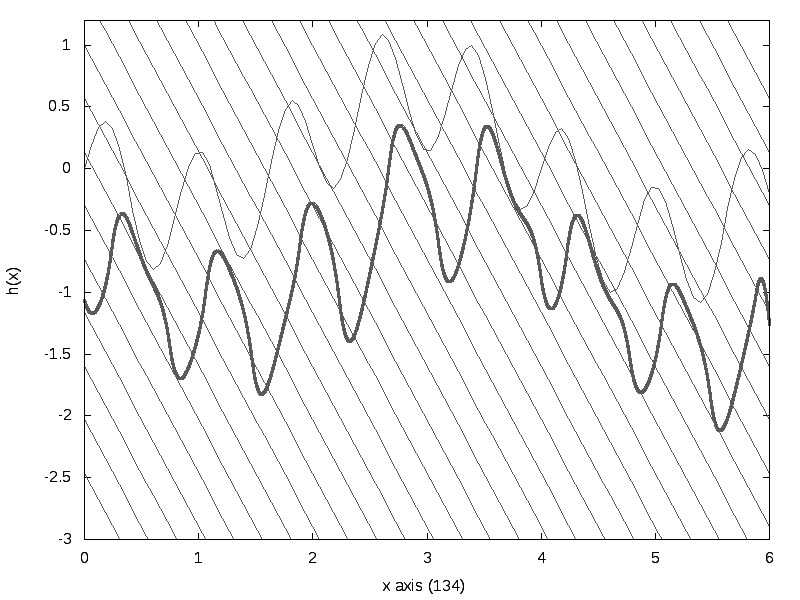}}
    \caption{Time evolution for $h(x)=\sin(\pi x)\cos(\frac{3\pi x}{2})+\frac{1}{5}\sin(\frac{\pi x}{4})$}
\end{figure}

In the second sequence, the computation is carried out for the same value of the constants. Here, another effect becomes manifest: first the peaks tilt and then, when the slope is aligned with the sunbeam, the lower zones start to undergo an enhanced sublimation rate.
Therefore, the differential equations adequately provide a positive feedback mechanism for the sublimation pits, as stated in previous literature \cite{Bergeron,Betterton,Lliboutry,Mat}, what is the ultimate cause behind the runaway process of penitente formation.

\section{Conclusions}

We have provided numerical evidence that the surface energy balance model proposed in Ref.\cite{Betterton}, accounts for the formation of spiky features in icy surfaces.

On the other hand, when one considers sunlight coming in at a given angle, the presence of shadows induces the tilting of the structures and a change in the shape of the crests.

Further considerations of vapor diffusion, and subsequent reabsorption, at the interface between the snow and the surrounding air, in a similar fashion as in Ref.\cite{Claudin}, can be introduced in order to improve the model.

\section*{Acknowledgments}

We thank M.G. Clerc, R. Rojas and R. Soto for interesting discussions. P.G. is a UNAB Ph. D. Scholarship Holder

\end{document}